\documentclass[12pt,a4paper]{article}
\pdfoutput=1
\usepackage{epsfig,amssymb,amsmath,graphicx,caption,subcaption,verbatim,hyperref,xcolor,ulem,epstopdf,psfrag,pstool,braket,array,enumerate,wrapfig,tabularx,wrapfig}
\usepackage{graphics,color, ltablex,booktabs}
\usepackage[font={small}]{caption}
\numberwithin{equation}{section}

\begin{document} 

\newcommand{\be}{\begin{equation}}
\newcommand{\ee}{\end{equation}}
\newcommand{\bea}{\begin{eqnarray}}
\newcommand{\eea}{\end{eqnarray}}
\newcommand{\bean}{\begin{eqnarray*}}
\newcommand{\eean}{\end{eqnarray*}}
\font\upright=cmu10 scaled\magstep1
\font\sans=cmss12
\newcommand{\ssf}{\sans}
\newcommand{\stroke}{\vrule height8pt width0.4pt depth-0.1pt}
\newcommand{\Z}{\hbox{\upright\rlap{\ssf Z}\kern 2.7pt {\ssf Z}}}
\newcommand{\ZZ}{\Z\hskip -10pt \Z_2}
\newcommand{\C}{{\rlap{\upright\rlap{C}\kern 3.8pt\stroke}\phantom{C}}}
\newcommand{\R}{\hbox{\upright\rlap{I}\kern 1.7pt R}}
\newcommand{\HH}{\hbox{\upright\rlap{I}\kern 1.7pt H}}
\newcommand{\CP}{\hbox{\C{\upright\rlap{I}\kern 1.5pt P}}}
\newcommand{\identity}{{\upright\rlap{1}\kern 2.0pt 1}}
\newcommand{\half}{\frac{1}{2}}
\newcommand{\quart}{\frac{1}{4}} 
\newcommand{\pr}{\partial}
\newcommand{\bm}{\boldmath}
\newcommand{\I}{{\cal I}} 
\newcommand{\M}{{\cal M}}
\newcommand{\N}{{\cal N}}
\newcommand{\e}{\varepsilon}
\newcommand{\ep}{\epsilon}
\newcommand{\balpha}{\mbox{\boldmath $\alpha$}}
\newcommand{\bgamma}{\mbox{\boldmath $\gamma$}}
\newcommand{\blambda}{\mbox{\boldmath $\lambda$}}
\newcommand{\bep}{\mbox{\boldmath $\varepsilon$}}
\newcommand{\Oh}{{\rm O}}
\newcommand{\x}{{\bf x}}
\newcommand{\y}{{\bf y}}
\newcommand{\bR}{{\bf R}}
\newcommand{\bl}{{\bf l}}
\newcommand{\bJ}{{\bf J}}
\newcommand{\X}{{\bf X}}
\newcommand{\Y}{{\bf Y}}
\newcommand{\z}{{\bar z}}
\newcommand{\w}{{\bar w}}
\newcommand{\tT}{{\tilde T}}
\newcommand{\tX}{{\tilde\X}}
\def\ir3{\int_{\mathbb{R}^{3}}}

\thispagestyle{empty}
\begin{center}
{{\bf \Large Evidence for Tetrahedral Structure of Calcium-40}} 
\\[15mm]

{\bf \large N.~S. Manton\footnote{email: N.S.Manton@damtp.cam.ac.uk}} \\[1pt]
%\vskip 1em
{\it 
Department of Applied Mathematics and Theoretical Physics,\\
University of Cambridge,
Wilberforce Road, Cambridge CB3 0WA, U.K.}
\vspace{1mm}

\end{center}

%%%%%%%%%%%%%%%%%%%%%%%%%%%%%%%%%%%%%%%%%%%%%%%%%%
\abstract{}
%%%%%%%%%%%%%%%%%%%%%%%%%%%%%%%%%%%%%%%%%%%%%%%%%%

More than 100 excited states of the Calcium-40 nucleus, with isospin 0, 
are classified into rotational-vibrational bands of an intrinsic 
tetrahedral structure. Almost all observed states below 8 MeV can be 
accommodated, as well as many high-spin states above 8 MeV. The bands 
have some similarity to those classifying states of Oxygen-16, but 
the A-mode vibrational frequency is lower relative to the E-mode and F-mode 
frequencies than in Oxygen-16. Previously identified rotational bands up 
to spin 16 and energy above 20 MeV are unified here into 
a smaller number of tetrahedral bands.    

\vskip 5pt

\vfill
\newpage
\setcounter{page}{1}
\renewcommand{\thefootnote}{\arabic{footnote}}

%%%%%%%%%%%%%%%%%%%%%%%%%%%%%%%%%%%%%%%%%%%%%%%%%%%%%%%%%%%%%%%%%%%%%%%%%%%%%
%%%%%%%%%%%%%%%%%%%%%%%%%%%%%%%%%%%%%%%%%%%%%%%%%%%%%%%%%%%%%%%%%%%%%%%%%%%%%

\section{Introduction} 

It was proposed nearly 100 years ago that the intrinsic structure of the
doubly magic Oxygen-16 nucleus is tetrahedral and that the nucleus is
a cluster of four alpha particles \cite{Wheel,Wef,HT}. The principal 
evidence is the presence of low-lying $J^P = 3^-$ and $J^P = 4^+$ 
states with their excitation energies having the approximate ratio 
$\frac{5}{3}$, as expected for a tetrahedral rotational band where the 
collective rotational energy scales as $J(J+1)$, and the absence of a 
$2^+$ state below the $3^-$ state. The strong transitions between the 
$3^-$ state and the $0^+$ ground state, and between the $4^+$ state 
and the ground state provide further evidence for collectivity. 
Interpreting the $3^-$ state as a 1-phonon, octahedral vibration of 
an intrinsically spherical nucleus is less convincing, because there 
is no obvious multiplet of close-to-degenerate 2-phonon states with 
spin/parities $6^+, 4^+, 2^+, 0^+$. 

Lezuo in the 1970s drew attention to similar features in the energy
spectrum of the two larger magic nuclei Calcium-40 and Lead-208 \cite{Lez}. 
Calcium-40 has a low-lying $3^-$ state, and then three $4^+$ states 
in the required energy range for a ratio $\frac{5}{3}$, but 
one has energy 15\% too low and the other two have energy 5\% too
high. Lezuo mentioned that a fuller understanding of the spectrum would
require a modelling of vibrational modes. This is what we initiate here.
In Lead-208, the yrast $4^+$ state has the right energy to be associated
with the well known yrast $3^-$ state, and Heusler has argued that eight
additional states appear to lie in collective tetrahedral bands \cite{Heu}.

The analysis of vibrations coupled to rotations has been quite
energetically pursued in the case of Oxygen-16. It was initiated 
by Wheeler \cite{Wheel} and by Dennison \cite{Den}, and followed up by Robson 
\cite{Rob}, Bijker and Iachello \cite{BI}, and others. 
The most detailed model was constructed recently by Halcrow, King and 
the present author \cite{HKM1,HKM2}. Our model was based on 
Skyrmion solutions \cite{Sky} in which alpha particles are extended 
structures of the Skyrme field, stabilised by their topological
charge (baryon number). Four of these extended alpha particles retain 
their identities and only slightly merge, forming a stable tetrahedral 
structure that is the basis for Oxygen-16 \cite{BMS}. This is not 
very different from the more familiar tetrahedral cluster of pointlike 
or spherical alpha particles.

A tetrahedron of alpha particles can vibrate and rotate. The vibrational 
modes of four particles connected by springs give a reasonable 
approximation to the vibrational modes of a tetrahedral Skyrmion with 
baryon number 16. There are A-, E- and F-mode vibrations: the A-mode 
is non-degenerate and preserves the tetrahedral symmetry, 
the E-mode is a doubly-degenerate quadrupole mode, and the F-mode 
is a triply-degenerate vector mode. Their frequencies (energies,
as we set $\hbar = 1$) are, respectively, approximately 12, 3.5 and 6 MeV.
The simplest quantised model of the dynamics has multiple phonon
states of these modes, coupled to rotations. These are known as 
rovibrational (rotational-vibrational) states. Each vibrational state
generates a rotational band, whose allowed spin/parities are well 
known in molecular physics \cite{Her}.

In Oxygen-16, the E-mode has lowest frequency, and in \cite{HKM1,HKM2}
the first excited $0^+$ state at 6.05 MeV is interpreted as a state 
with two E-phonons. The E-mode, if extended to large amplitude, can deform
a tetrahedron of four alpha particles through a square configuration 
into the dual tetrahedron (the spatial inversion of the initial 
tetrahedron). As the intermediate square configuration has higher energy, 
this is quantum mechanically a tunnelling process. Its importance 
has been long recognised \cite{Den}. Without tunnelling, the Oxygen-16
spectrum would have many parity doublets, but the tunnelling splits
these, matching the data. In \cite{HKM1} we constructed a two-dimensional 
E-manifold, which extended the space of E-mode vibrations
so as to include tetrahedral configurations of both types and square 
configurations. Quantum mechanics on this E-manifold replaces the 
harmonic oscillator quantisation of the E-mode, but states can still 
be classified in terms of E-phonons (by examining the wavefunctions 
close to the initial tetrahedral configuration). The contributions of 
A-phonons and F-phonons were added in \cite{HKM2}; these were treated 
harmonically.

By combining E-manifold states with A- and F-phonons, and rotations,
we obtained a spectrum of isospin 0 states of Oxygen-16, classified 
by their spin/parity and energy, matching just about every known state of
Oxygen-16. The matching is precise up to 15 MeV above the ground state
-- there is just one predicted $4^-$ state with energy 13 -- 14 MeV that
has not been observed. Above 15 MeV, more states are predicted than
have been observed, but the experiments are rather incomplete here,
and the analysis difficult. For example, our theory predicts a few
$6^-$ states between 15 and 20 MeV, but none has yet been
observed, partly because such states are inaccessible to some of the 
experiments. Several $6^+$ states are known in this range, and it is
likely that $6^-$ states are present too. Between 20 and 30 MeV several
high-spin states have been seen, up to spin 9 and maybe spin 10. These
are also predicted by our model, but the match to energies, and even
to the number of states, is not very good.

One surprise in all this is that there appears no need to identify any states
as shell-model excitations. They are all collective. 
Alternative descriptions of the excited states
of Oxygen-16 are available, entirely in terms of the shell model, see e.g.
\cite{TWC}, p.7. The conclusion seems to be that collective
excitations and shell-model excitations are alternative ways 
of modelling the same nuclear states, and 
it is not necessary to treat them as independent. The reason seems to
be the strong mixing that occurs between pure shell-model excitations. 
This applies even for the lowest-energy 1-particle, 1-hole (1p1h)
excitations. By contrast, in the isospin 1 sector of Nitrogen-16, 
Oxygen-16 and Fluorine-16, the lowest states are clearly a quartet of 
simple 1p1h excitations, with very similar energies. The particles
are in the ${\rm d}_{\frac{5}{2}}$ and ${\rm s}_{\frac{1}{2}}$ states of
the sd-shell, and the holes are in the ${\rm p}_{\frac{1}{2}}$ states
of the p-shell. Combined, they give states with spin/parities
$3^-, 2^-, 1^-, 0^-$ \cite{Suh}. States with the same spin/parities should 
also occur with isospin 0 in Oxygen-16, and they do, as yrast states, but 
with a greater spread of energies. However, these states are also easily 
identifiable with collective states, contributing to rotational
bands. For example, the lowest $1^-$ state is the bandhead of the F-band, 
with one F-phonon. If this state were identified purely as a 
shell-model state, it would spoil the interpretation of several 
higher-spin, higher-energy states as lying in this F-band. The only
state whose 1p1h character appears to have a dominant influence is the yrast
$0^-$ state. In the collective interpretation, this state has three
E-phonons, and a rather complicated E-manifold wavefunction with rather
high energy. The overlap with a 1p1h wavefunction should reduce 
its energy closer to the observed value.

Let us turn now to Calcium-40. Finding a Skyrmion with baryon number 
40 to model Calcium-40 has been difficult, but using a suitable 
multi-layer structure as starting point for a numerical relaxation, 
a tetrahedrally symmetric solution was found \cite{Lau}. This solution 
is shown in Figure 1 in two orientations. It resembles the tetrahedral 
Skyrmion of baryon number 56 \cite{LM} (56 is a tetrahedral number), 
with four clusters of baryon number 4 chopped off at the vertices. 
The final structure is closer to spherical, without the pointedness of 
an ideal tetrahedron, and therefore more stable. 

\begin{figure}
\begin{minipage}[c]{0.4\linewidth}
\includegraphics[width=\linewidth]{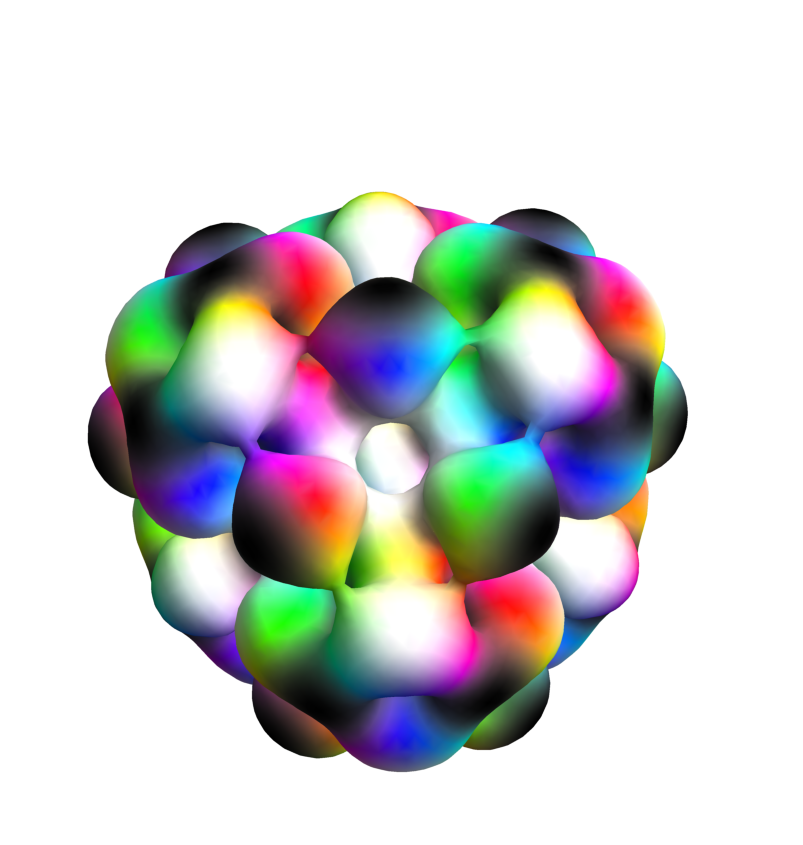}
\end{minipage}
\hfill
\begin{minipage}[c]{0.4\linewidth}
\includegraphics[width=\linewidth]{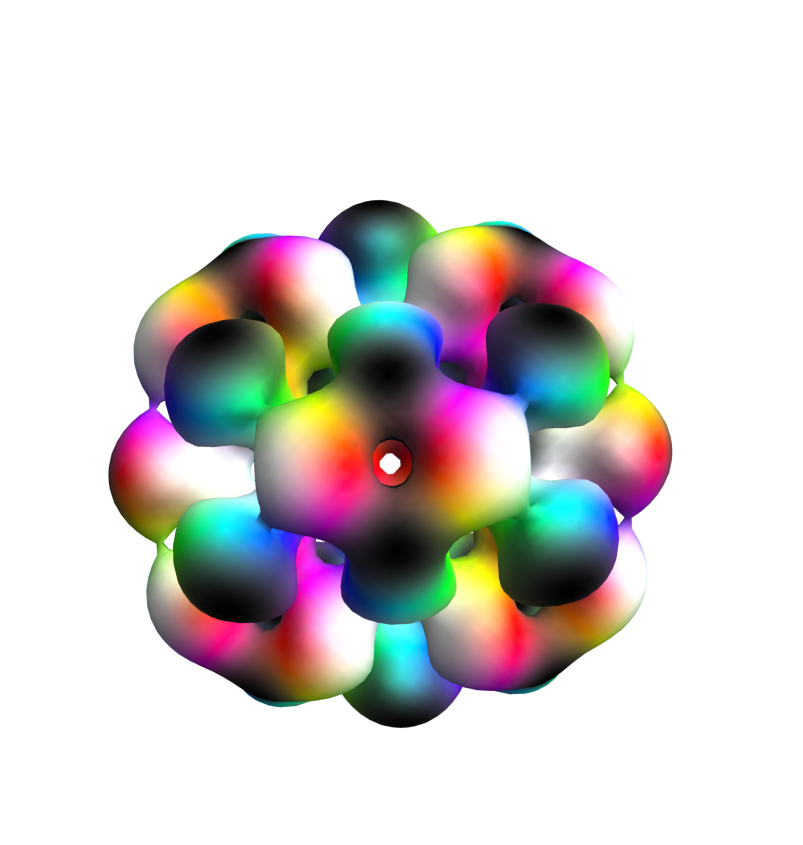}
\end{minipage}
\caption{Skyrmion with baryon number 40 in two orientations
  (figures courtesy P.H.C. Lau).}
\end{figure}

This Skyrmion solution also emerges by stacking 40 basic Skyrmions of
baryon number 1 in a cluster with tetrahedral symmetry. A structure
like this arises as a subcluster of the FCC lattice, or as a weight
diagram of a representation of SU(4) (all SU(4) weight diagrams have 
at least tetrahedral symmetry) \cite{Man,HMR}. A model of the relevant cluster 
is shown in Figure 2, in the same orientations as for the Skyrmion. 
The steel balls correspond to the cores of the basic Skyrmions, and the
magnetic rods connecting them correspond to the nearest-neighbour attractive 
forces -- these are not materially present in the Skyrmion solution. 
Physically, such a cluster appears as a static solution in a variant of the 
Skyrme model called the lightly bound model, where the basic Skyrmions 
attract but hardly merge \cite{GHS,GHK}. It is particularly clear in
the lightly bound model how to orient the basic Skyrmions to achieve 
maximal attraction and greatest stability. The Skyrmion in Figure 1, 
which is a solution of the standard Skyrme model (with sigma model
term, Skyrme term, and pion mass term in the Lagrangian), can be 
obtained by relaxation of its analogue in the lightly bound model.

\begin{figure}
\begin{minipage}[c]{0.4\linewidth}
\includegraphics[width=\linewidth]{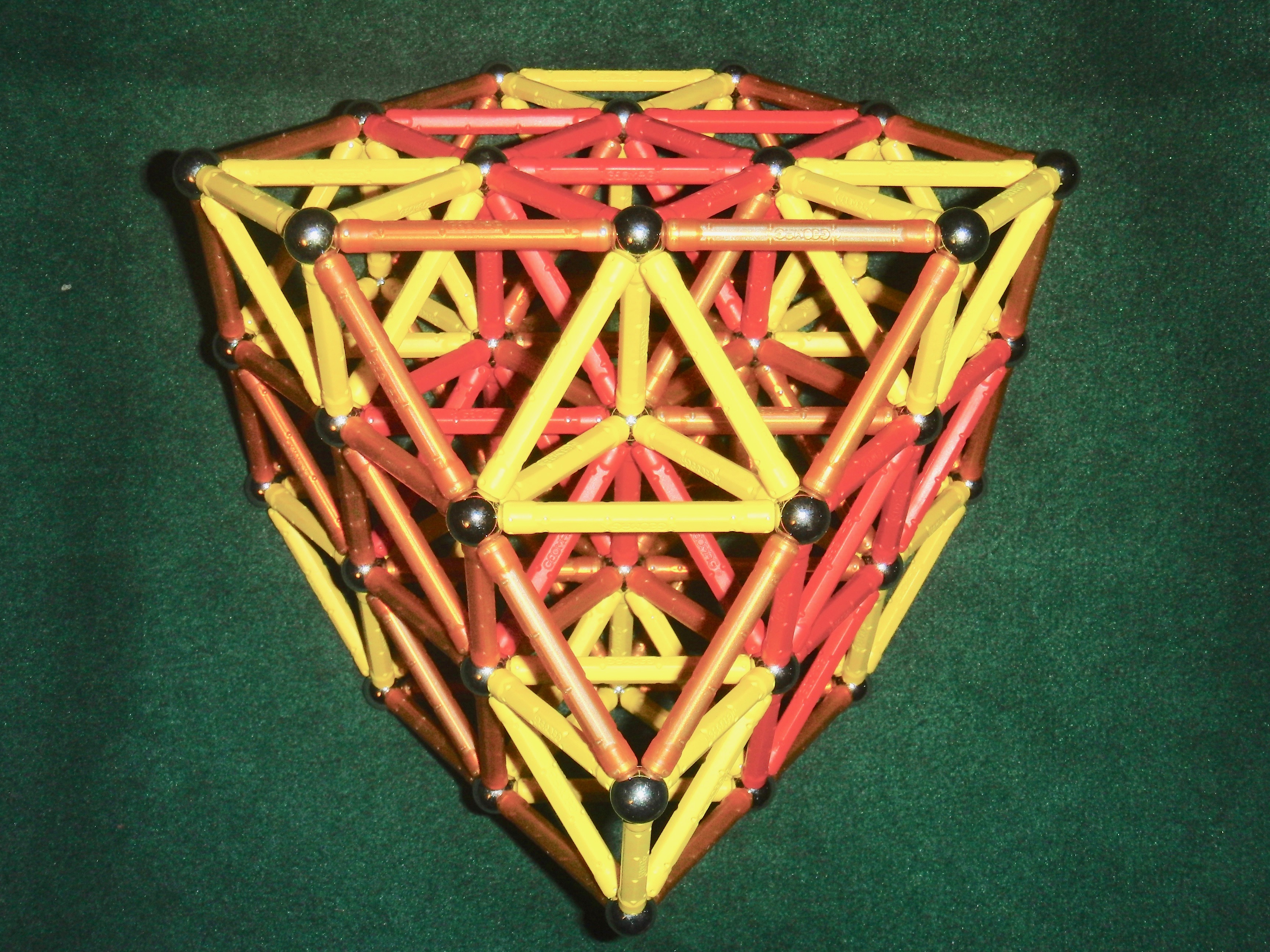}
\end{minipage}
\hfill
\begin{minipage}[c]{0.4\linewidth}
\includegraphics[width=\linewidth]{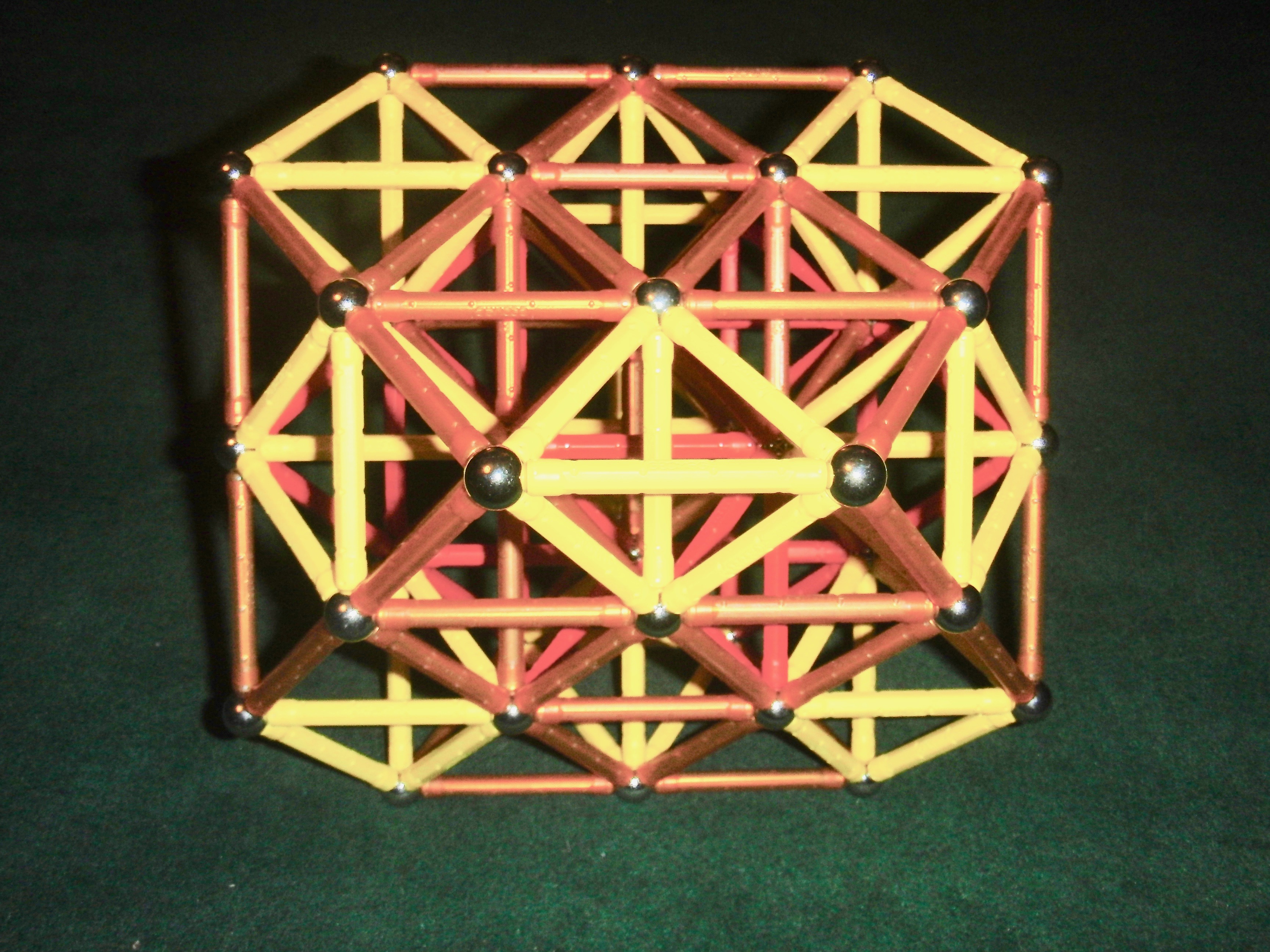}
\end{minipage}
\caption{Model cluster with baryon number 40 in two orientations.}
\end{figure}

Notice that the Skyrmion in Figure 1 has the property that it can be
identified as a bound cluster of 10 distinct tetrahedral Skyrmions of baryon 
number 4, i.e. Skyrme model alpha particles. These tetrahedral 
subclusters are highlighted by yellow magnetic rods in Figure 2.

We will not use the Skyrmion in our later discussion, but it is 
encouraging that a tetrahedrally symmetric, quasi-spherical Skyrmion 
with baryon number 40 exists. Instead, we will present a phenomenological 
model of states of Calcium-40, based on the rovibrational spectrum of a 
tetrahedron. This will involve A-, E- and F-phonons, combined with 
rotations. The model here is less detailed than the model for Oxygen-16 
in \cite{HKM2}, but qualitatively similar.

One significant difference, compared with Oxygen-16, is that states
of Calcium-40 have been classified up to spin 16, whereas for
Oxygen-16 the highest confirmed spin is about 9. Such high spins have
not been measured directly, but inferred from gamma-ray spectra. We
make substantial use of the interesting and detailed spectra of Calcium-40 
states with positive parity, up to $16^+$, that have been tabulated and 
gathered into proposed rotational bands in \cite{Ide}. There, no use 
is made of the theory of tetrahedral vibrations. The negative 
parity states are less completely classified, but we use 
the spectra tabulated and gathered into bands in \cite{Tor}. There again,
a tetrahedral structure is not discussed, but it is argued that 
some type of permanent octupole deformation needs to be considered.   
For the complete set of experimentally determined states we use the 
standard compilation \cite{ENSDF}.

Perhaps our most important observation, providing evidence for the
tetrahedral structure of Calcium-40, is that the number of states in
two of the rotational bands with even spin and positive parity, described 
in \cite{Ide}, jumps by one starting at $8^+$ and jumps by 
one again starting at $14^+$. This is just as expected for the single
rotational band of a tetrahedron excited by one E-phonon, 
in contrast with rotational bands of an ellipsoidally deformed
nucleus, where there are no such jumps. More generally, we show that 
several of the rotational bands identified in \cite{Ide} and
\cite{Tor} are naturally unified into far fewer bands when interpreted as
rovibrational excitations of a tetrahedron. 

Four states of Calcium-40 can be clearly identified as 1p1h excitations. 
The particle states are from the ${\rm f}_{\frac{7}{2}}$ shell and the 
holes are ${\rm d}_{\frac{3}{2}}$ states from the sd-shell. Combined, they 
give states with spin/parities $5^-, 4^-, 3^-, 2^-$. Their energies, and
even their ordering, are not easily or precisely predicted by shell-model
calculations \cite{Suh}, but the trend is for the energy to decrease as
the spin increases. In particular, the $5^-$ state, identified with
the observed yrast $5^-$ state at 4.491 MeV, is strikingly low in energy.

These four 1p1h states of Calcium-40 are not easily fitted into
collective rotational bands. However, our model for Oxygen-16 suggests
that they do not need to be treated as
additional to the collective states. We will show later that it is
best to identify these four states as belonging to rotational bands,
but with significantly displaced energies. As collective states, they 
have wavefunctions and energies that are significantly affected by their 
overlap (superposition) with pure 1p1h excitations, but we do not need to 
postulate that they are totally distinct. The effect is largest for 
the $5^-$ state. In our model it lies in the E-band, with one
E-phonon, but its energy is shifted down by more than 2 MeV compared
to what would be expected by extrapolation from other states in the E-band.

\section{Rovibrational bands of a tetrahedron}

Here we review the allowed spin/parities of rovibrational states of a 
vibrationally excited tetrahedron. They are obtained by using the character
table of the tetrahedral group $T_d$ \cite{LL}, which is reproduced as
Table 1 in Appendix A. Spins up to 12 are tabulated in 
\cite{Her}, p.450 but we need some results up to spin 16.

The group $T_d$ has five irreducible representations (irreps), denoted
${\rm A}_1$, ${\rm A}_2$, ${\rm E}$, ${\rm F}_2$, ${\rm F}_1$, where 
${\rm A}_1$ (the trivial irrep) and ${\rm A}_2$ are 1-dimensional, 
${\rm E}$ is 2-dimensional, and ${\rm F}_2$ and ${\rm F}_1$ are 
3-dimensional. It appears that, as for Oxygen-16, the dominant 
vibrational modes of the intrinsic tetrahedron modelling Calcium-40 
are A-, E- and F-modes, transforming according to the ${\rm A}_1$, ${\rm E}$ 
and ${\rm F}_2$ irreps. 1-phonon states are classified by these three irreps.

Multiphonon states also occur. As phonons are bosons, multiphonon 
states transform as a symmetrised tensor product of 1-phonon irreps. 
These reducible representations of $T_d$ can then be decomposed into 
irreps. Particularly important for us is that the space of vibrational states 
with two E-phonons decomposes as
\be  
{\rm E}^2 = {\rm A}_1 \oplus {\rm E} \,.
\label{E2decomp}
\ee
(For a representation $d$ with character $\chi$, the symmetric square
$d^2$ -- we drop the subscript ``symm.'' -- has character 
$[\chi^2](g) = \half\{\chi(g)^2 + \chi(g^2)\}$ for 
each group element $g$.) See \cite{HKM2} or \cite{Her} p.127 
for higher symmetrised powers of ${\rm E}$.

To find the allowed spin/parities of the rotational excitations of a
given vibrational state, one starts with the characters $\chi$ of the
spin $J$ representation of the group O(3) with positive or negative 
parity ($J$ is always an integer here). For a pure rotation by angle 
$\theta$, the character is 
\be
\chi(\theta) = \frac{\sin(J+ \half)\theta}{\sin\half\theta}
\label{charrot}
\ee
for either parity. For a rotation by $\theta$ combined with an 
inversion $I$ it is
\be
\chi(\theta, I) = \pm \, \frac{\sin(J+ \half)\theta}{\sin\half\theta}
\label{charrotinv}
\ee
with the upper (lower) sign for positive (negative) parity. Next, 
these spin $J$ characters need to be restricted to the elements of the group
$T_d$, which lie in five conjugacy classes -- rotations by $0$,
$\frac{2\pi}{3}$ and $\pi$, and rotations by $\pi$ and $\frac{\pi}{2}$
combined with inversion. The characters of the spin $J$
representation of O(3) thereby become characters of a (generally 
reducible) representation of $T_d$, and using the character table one 
can find its decomposition into irreps. If an irrep $d$ occurs here 
and simultaneously as an irrep of a multiphonon excitation (as for example,
on the right hand side of (\ref{E2decomp})), then that multiphonon 
excitation can have a rovibrational state of spin $J$ and the
appropriate parity. If $d$ occurs with multiplicity $n$ then the 
spin $J$ state occurs with multiplicity $n$. 

For example, the spin 2, positive parity representation of O(3) is 
5-dimensional and its characters for the five tetrahedral conjugacy 
classes are, respectively, $5, -1, 1, 1, -1$. Its decomposition into 
irreps of $T_d$ is therefore
\be
(J^P = 2^+)\Bigl\arrowvert_{T_d} = {\rm E} \oplus {\rm F}_2 \,,
\label{2+decomp}
\ee
so $2^+$ states occur in the rovibrational E-band (one
E-phonon) and in the F-band (one F-phonon), and also in multiphonon 
rovibrational bands where the multiphonon state has an irrep 
${\rm E}$ or ${\rm F}_2$ in its $T_d$ decomposition.

Using the formulae (\ref{charrot}) and (\ref{charrotinv}), and 
the $T_d$ character table, one finds that the ground state band, the
rotational excitations of a tetrahedron with no vibrational phonons,
has states with spin/parities 
\bea
{\rm g.s. band}: && 0^+, 3^-, 4^+, 6^+, 6^-, 7^-, 8^+, 9^+, 9^-, 
10^+, 10^-, 11^-, 12^+, 12^+, 12^-, \nonumber \\ 
&& 13^+, 13^-, 14^+, 14^-, 15^+, 15^-, 15^-, 16^+, 16^+,
16^-, \dots \,.
\label{Alist}
\eea 
(Note that multiplicities greater than 1 start at spin 12.)
There is some periodic structure here. Whenever the spin increases 
by 6, the number of allowed states (ignoring the parity label) increases by
1. So, for example, there are no spin 2 states, one spin 8 state and
two spin 14 states. This is because the characters of O(3), for
rotations by $\frac{2\pi}{3}$ and $\pi$, repeat when $J$ increases by
6, whereas the character for the trivial rotation by $0$, which is $2J+1$, 
increases linearly with $J$. Similarly, the difference between the number of
positive parity and negative parity states has period 4 as $J$ increases, 
being $1,0,0,-1$ for $J = 0,1,2,3 \, \mod 4$. This is because the characters 
for the rotations by $\pi$ and $\frac{\pi}{2}$, combined with 
inversion, repeat when $J$ increases by 4.

There is another feature of the states in the ground state band (\ref{Alist}). 
The set of allowed spins (ignoring the parity labels and the multiplicities) 
forms a numerical semigroup \cite{NumSem}. This is explained in Appendix B. 

The next important band is the E-band, the rovibrational states where the
tetrahedron is excited by one E-phonon. Here the states
all occur symmetrically as parity doubles. The allowed spin/parities are 
\bea
\textrm{E-band}: && 2^\pm, 4^\pm, 5^\pm, 6^\pm, 7^\pm, 8^\pm, 8^\pm, 
9^\pm, 10^\pm, 10^\pm, 11^\pm, 11^\pm, 12^\pm, 12^\pm, \nonumber \\
&& 13^\pm, 13^\pm, 14^\pm, 14^\pm, 14^\pm, 15^\pm, 15^\pm, 16^\pm, 
16^\pm, 16^\pm, \dots \,.
\label{Elist}
\eea
In the E-band, the number of states of each parity increases by 1 whenever 
the spin $J$ increases by 6. In particular, there is an upward step in
the number of allowed states at spin 8 and spin 14, and these steps are
observed experimentally \cite{Ide}. For low spins, the 
energies are higher in the E-band than in the ground
state band because of the E-phonon energy, but the bands cross over at
higher spin, because the phonon stretches the nucleus, increasing its
effective moment of inertia. 

A further band is the 2-phonon, ${\rm E}^2$-band. As pointed out in 
eq.(\ref{E2decomp}), ${\rm E}^2 = {\rm A}_1 \oplus {\rm E}$. The ${\rm A}_1$ 
irrep is trivial, so 2-phonon states transforming under this 
irrep have rotational excitations with the same spin/parities 
as in the ground state band, (\ref{Alist}). 2-phonon states transforming 
under the E irrep have rotational excitations with spin/parities 
as in the E-band, (\ref{Elist}). The energies are slightly higher for
the 2-phonon than 1-phonon states, and the E-band and ${\rm E}^2$-band
appear not to cross over.

The triply-degenerate vibrational states with one F-phonon have a vectorial
character, and carry an internal spin/parity $1^-$. Their rotational 
excitations combine a spin/parity $J^P$ from the ground state band 
(\ref{Alist}) with the internal spin/parity $1^-$, using the usual 
Clebsch--Gordon rules. This gives the F-band spin/parities
\be
\textrm{F-band}: \ \ 1^-; 2^+, 3^+, 4^+; 3^-, 4^-, 5^-; 5^\pm, 6^\pm, 
7^\pm; \dots \,,
\label{Flist}
\ee
where the states are grouped according to their ground state band
spin/parity, i.e. the values $0^+, 3^-, 4^+, 6^\pm, \dots$. Because of 
the relatively high energy of the F-phonon, we do not need to 
consider higher spins here.

\section{Interpreting the Calcium-40 spectrum}

As primary data for the spectrum of excited states for Calcium-40
we use the ENSDF adopted levels \cite{ENSDF}. We have assigned 104 of
these states to tetrahedral rovibrational bands. The assignments are
listed in Table 2 of Appendix C, and the energies and spins of the states
are shown in Figure 3. Filled circles (triangles) indicate positive 
(negative) parity states, and the colouring distinguishes the various
rovibrational bands. States with a given spin, in one band, do not all 
have the same energy. We saw this in Oxygen-16, where the energy
splittings could be calculated \cite{HKM1,HKM2}. Part of the splitting between 
positive and negative parity states arose from the tunnelling 
between a tetrahedral configuration and its dual. We have not made new 
calculations for the splittings here, as we do not have a precise
dynamical model for the tetrahedral deformations.

\begin{figure}
\vskip -100pt
\begin{center}
\includegraphics[width=1\textwidth]{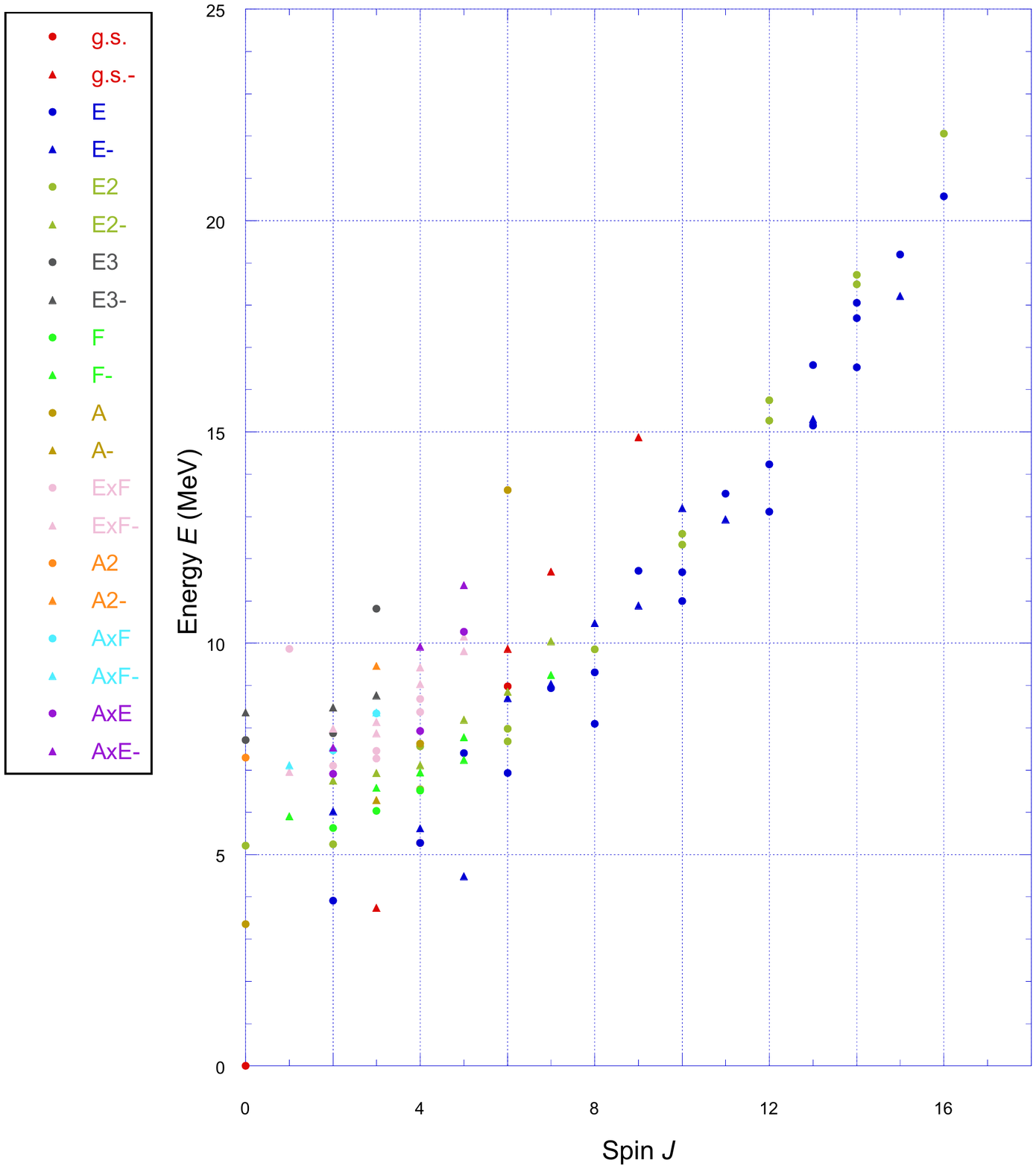} 
\end{center}
\vskip -60pt
\caption{Observed Calcium-40 energy spectrum, with the rovibrational
band assignments in our model. Positive parity states are displayed 
as filled circles, negative parity states as filled triangles. 
(E2- denotes the negative parity part of the ${\rm E}^2$-band, etc.)} 
\label{spectrum}	
\end{figure}

From \cite{ENSDF} we retain almost all states below 8 MeV. The states 
at 6.160, 6.422, 7.421 and 7.481 MeV are dropped as they are either 
doubtful or have unknown (low) spins. The states at 
7.658 and 7.694 MeV are dropped as they have isospin 1.
Above 8 MeV we retain selected states, mainly of spin 5 and higher. 
States of lower spin are numerous and would lie in multiphonon bands, 
and we have only attempted to classify a few of these. In several cases, in the
range between approximately 7 and 10 MeV, the experiments only partially 
constrain the possible spin and parity of a state. In these cases, we 
have made a choice that seems most consistent with the pattern of 
bands that we have partially determined using other states. These
choices are shown in Table 2 and in Figure 3. Most of 
them are for states classified as having two phonons. 

Above 9 MeV, high-spin states have their energy and spin 
mainly inferred from the pattern of gamma decays, rather than direct 
measurement, and we retain all the states that have been assigned to bands 
in \cite{Ide} and \cite{Tor}. The highest of these is a $16^+$ state 
at 22.106 MeV. 

The band assignment that we are most confident about is the E-band. We
can identify 31 observed states up to spin 16, of positive and 
negative parity, as being in this band. They are shown in blue in Figure 3. 
Many are from the positive parity bands 2, 3 and 4 as identified 
in \cite{Ide}, and from the negative parity bands in \cite{Tor} 
(rightmost two columns in Fig. 5). Just a few expected states of 
positive parity are missing from this E-band -- a second $11^+$ state, 
a second $15^+$ state, and a second and third $16^+$ state are
expected, according to the list (\ref{Elist}), but have
not yet been identified. A greater number of negative parity states are 
missing. This is because there is just one identified state with each 
of the spin/parities $8^-, 10^-, 11^-, 13^-, 15^-$ in \cite{Tor}, 
but the E-band requires two. Of more concern is that two $12^-$ states, and 
three $14^-$ and $16^-$ states are expected, but none of these have 
been observed or inferred. Maybe a reanalysis of the data, taking
these expectations into account, would reveal the missing states.  

The plot of energy $E$ against spin $J$ is approximately linear for the
E-band. It is nothing like a linear relation between $E$ and
$J(J+1)$. Presumably the E-phonon softens the nucleus, and centrifugal
stretching produces an effective moment of inertia that increases 
with $J$. The approximate linearity with $J$ allows us to extrapolate the
E-band down in energy towards spin 0. In this way we have identified 
the $2^\pm$ and $4^\pm$ states to be assigned to the E-band. We can
also read off the E-phonon frequency to be approximately 2.5 MeV.

The negative and positive parity states are fairly well interleaved in
the E-band. An outlier, below the general trend, is the yrast $5^-$
state at 4.491 MeV. This is interpreted as in the E-band but with lowered
energy because of its partial 1p1h character. The $4^-$ and $2^-$
states in the E-band are also yrast states, and are interpreted as 
having a partial 1p1h character, but a less pronounced lowering of 
energy. The final 1p1h excitation, with spin/parity $3^-$ cannot be in 
the E-band at all. It probably contributes to the yrast $3^-$ state at
3.737 MeV, which is assigned here to the ground state band.

Unifying several of the rotational bands from refs.\cite{Ide,Tor} into one
tetrahedral rovibrational band is theoretically attractive, not only 
because it explains some of the observed jumps of multiplicity as the 
spin increases. It also appears to be supported by the many 
cross-band gamma transitions that are observed. For example, 
transitions of roughly equal strength connect the multiple states 
with spin above 10 (see Fig. 1 of \cite{Ide}).
 
Only a few states can be identified as being in the ground state band. 
It appears that here the intrinsic shape is more rigid, and that the energy 
$E$ scales more closely with $J(J+1)$. The $0^+$ ground state and $3^-$
yrast state at 3.737 MeV are clearly in this band, but we suggest that
above the $3^-$ state the ground state band crosses the
E-band, so that the $6^\pm$, $7^-$ and $9^-$ states in the ground state
band are well above those in the E-band. (Our assignment of the $6^-$
state at 9.860 MeV to the ground state band is tentative.)  
This band crossing is similar to what occurs in Oxygen-16, where 
the ground state band crosses the E-band between spin 4 
and spin 6 \cite{Rob,HKM2}.

The $4^+$ state at energy 5.279 MeV is a mystery. It occurs at the
crossover of the ground state band and the E-band. Uniquely, we have 
assigned it to both these bands. Really, two independent states are
needed for these bands. Possibly the observed state at 6.160 MeV,
which has an uncertain $3^-$ spin/parity assignment, and which we have
currently omitted from our bands, is the further $4^+$ state we need. 
Alternatively, we have misassigned the low-lying $4^+$ states. 
In any case, between 5 and 7 MeV there are apparently just three $4^+$ 
states observed, but our band structure needs four.  

The E-band does not accommodate all the known positive parity states with
high spin. In particular, higher-lying observed states with
spin/parities $8^+, 10^+, 12^+, 14^+, 16^+$ are in a separate
band, shown as band 1 in \cite{Ide}. We assign them to the
${\rm E}^2$-band. As noted before in
eq.(\ref{E2decomp}), the ${\rm E}^2$-band splits into an ${\rm A}_1$ part 
and an ${\rm E}$ part. In modelling Oxygen-16 we discovered that the 
${\rm A}_1$ part has clearly lower energy, and we assume the same 
holds for Calcium-40. So the high-spin states listed above are in a band 
with the same spin/parities as in the ground state band. We 
assume this is again a rather soft band, with an approximately 
linear relationship between $E$ and $J$. The energies in this 
${\rm A}_1$ part of the ${\rm E}^2$-band 
are lower than those of the ground state band for spins down from 16 
to 6, but then the bands cross. By extrapolating 
down to spin 0, we identify the second excited $0^+$ 
state at 5.212 MeV as the ${\rm E}^2$-bandhead. Its energy 
matches our estimate for the energy of two E-phonons rather well. 

The high-spin states in the ${\rm E}$ part of the ${\rm E}^2$-band
have not been identified -- they would include, for example, further
$14^+$ states. But a few low-spin states in this part of the band can be
identified, in particular the states with spin/parity $2^\pm$. 

We need an interpretation for the first excited $0^+$ state at energy
3.353 MeV. This is assigned as the A-bandhead, a state with one 
A-phonon. The A-phonon energy is therefore approximately 3.4 MeV, 
somewhat higher than the E-phonon energy. The A-band is assumed 
to be less soft than the bands with one or two E-phonons, because the
nuclear shape stays closer to spherical. Like the ${\rm A}_1$ part of the
${\rm E}^2$-band, the A-band has the spin/parities of the ground state 
band, and we can identify its $3^-$ and $4^+$ states, and possibly its
$6^+$ state, all having energies about 3 MeV higher than those in the 
ground state band. However, there is a shortage of observed $6^+$ states 
in the required energy range, and a worse shortage of $6^-$ states.

A further band that we recognise is the F-band, with spin/parities as
in (\ref{Flist}). This includes the $1^-$ and $3^+$ yrast states at
5.903 and 6.030 MeV, which cannot occur in other bands. Using the 
energies of these two states we estimate the F-phonon energy to be 
about 5 MeV, higher than the E-phonon and A-phonon energies. 
We have identified states in the F-band up to spin 5 and just beyond, 
but with considerable uncertainty. 

Above the F-band we find evidence for an 
${\rm A} \times {\rm E}$ band, an ${\rm A} \times {\rm F}$ band, and an 
${\rm E} \times {\rm F}$ band. The allowed spin/parities in the 
${\rm A} \times {\rm E}$ band are the same as in the E-band, starting with
$2^\pm$ and $4^\pm$, and those in the ${\rm A} \times {\rm F}$ band
are as in the F-band, starting with $1^-$ and $2^+$. Those in the 
${\rm E} \times {\rm F}$ band are parity doubles of the states in 
the F-band, as ${\rm E} \times {\rm F} = {\rm F}_1 \oplus {\rm F}_2$, 
so the first few are $1^\pm$, $2^\pm$, and then $3^\pm$ with
multiplicity two. The energies of all these states are estimated to 
be the sum of the energies of the two contributing phonons and the
spin energy. (Recall that the A-, E- and F-phonons have approximate
energies 3.4, 2.5 and 5 MeV, respectively.) 
Suitable states to place in these bands can be tentatively 
assigned, by making appropriate choices among the spin/parity 
assignments in \cite{ENSDF}. These choices are shown in Table 2 and 
Figure 3. Rather a lot of choices are needed, so some are likely to 
be incorrect. A few further states with low spins are tentatively assigned to 
other multiphonon bands, including bands with two A-phonons, three 
E-phonons, or two E-phonons combined with one A-phonon (this last band
is omitted from Figure 3). The band with three E-phonons is the 
lowest that can accommodate a $0^-$ state.

\section{Conclusions}

We have shown that more than 100 observed states of Calcium-40, with 
isospin 0, can be assigned to rovibrational bands of an intrinsically 
tetrahedral nuclear structure. This makes Calcium-40 similar to the
smaller doubly magic nucleus Oxygen-16. Vibrational A-, E- and 
F-modes having approximate frequencies 3.4, 2.5 and 5 MeV can be 
excited, and we have identified states with up to three vibrational 
phonons. In particular, 31 states are assigned to the band with one 
E-phonon. Relative to the E-mode and F-mode frequencies, the A-mode 
frequency appears to be considerably lower than it is in 
Oxygen-16. Possibly this is because in Oxygen-16 the A-mode is a true 
breathing mode of the four alpha particles, whereas in Calcium-40 it 
is more likely to be a tetrahedrally symmetric, volume-preserving 
vibration, where six alpha particles move outward and four move inward.  

Our scheme accommodates almost all observed states with excitation 
energy below 8 MeV, as well as all the known states of higher energy 
with spins from 8 up to 16. It explains why there are few observed 
$0^-$, $1^+$ and $3^+$ states -- these can only occur in multiphonon 
bands (with the exception of one $3^+$ state in the F-band). It also 
explains why $2^+$ states are more numerous than $2^-$ states -- the 
F-band and ${\rm A} \times {\rm F}$ bands exclude the latter. Similarly, 
$4^+$ states are more numerous than $4^-$ states because the ground 
state band and A-band exclude $4^-$ states.

The scheme has a few gaps -- one more $4^+$ state is
predicted between 5 and 7 MeV, and a few extra states of spin 5 and 
spin 6 are expected between 8 and 11 MeV -- these have not yet been 
observed. A number of extra states with spin 8 or more, particularly 
of negative parity, are also expected from 11 MeV upwards. One of
these is the $8^+$ state in the ground state band, with a predicted
energy between 12 and 13 MeV.

The scheme could be made more robust if the spins and parities of
20 states between about 7 and 10 MeV were better known. At present, 
it is necessary to make choices among the experimentally allowed values
in order to fill some of the higher bands, but these choices cannot 
be regarded as firm predictions, as there are too many of them.

On the theoretical side, it would be helpful to have more information
about the Skyrmion with baryon number 40, especially its low-lying
vibrational modes. More generally, it would be helpful to investigate the
dynamics of clustering and vibrations in models with 10 alpha particles.

%%%%%%%%% Acknowledgements %%%%%%%%%%%%%
\section*{Acknowledgements}
%%%%%%%%%%%%%%%%%%%%%%%%%%%%%%%%%%%%%%%%

I am grateful to Anneli Aitta for help with tabulating the data and 
producing Figures 2 and 3. I thank Andreas Heusler for discussions and
Chris Halcrow for comments. This work has been partially supported by 
STFC consolidated grant ST/P000681/1.

%\vfill
%\newpage

\section*{Appendix A: Character table of $T_d$}

The 24 elements of $T_d$ lie in five conjugacy classes, denoted $Id$,
$C_3$, $C_2$, $\sigma_d$ and $S_4$. These consist, respectively, of 
rotations by $0$, ${\frac{2\pi}{3}}$ and $\pi$, and rotations by
$\pi$ and $\frac{\pi}{2}$ combined with an inversion.
The character table for the five irreps is \cite{LL}

\begin{tabularx}{\linewidth}{ l | l l l l l} \hline
  & $Id$ & $C_3$ & $C_2$ & $\sigma_d$ & $S_4$ \\ \hline
${\rm A}_1$ & 1 & 1 & 1 & 1 & 1 \\
${\rm A}_2$ & 1 & 1 & 1 & -1 & -1 \\
${\rm E}$ & 2 & -1 & 2 & 0 & 0 \\
${\rm F}_2$ & 3 & 0 & -1 & 1 & -1 \\
${\rm F}_1$ & 3 & 0 & -1 & -1 & 1 \\

\hline
		
\caption{Character table of $T_d$.}

\end{tabularx}

\renewcommand{\theequation}{B.\arabic{equation}}
\setcounter{equation}{0} 

\section*{Appendix B: Numerical semigroups}

It is noteworthy that in the ground state rotational band, the spins
that occur form a {\it numerical semigroup} \cite{NumSem}. This is the case not
just for an object whose intrinsic structure is tetrahedral, with
symmetry $T_d$, but for any other symmetry. The allowed
spins for, respectively, objects with tetrahedral, cubic and
icosahedral symmetries, i.e. $T_d$-symmetry, $O_h$-symmetry
and $Y_h$-symmetry, are \cite{BJ}
\bea
T_d : && 0,3,4,6,\dots \,, \\
O_h : && 0,4,6,8,9,10,12,\dots \,, \\
Y_h : && 0,6,10,12,15,16,18,20,21,22,24,25,26,27,28,30,\dots \,.
\eea
In each case, the dots $\dots$ indicate that {\it all} integers beyond the
last one shown are present.

A numerical semigroup $S$ is defined as a subset of the non-negative 
integers that includes 0 and is closed under addition; further, the 
set $G(S)$ of missing integers -- the gaps -- is finite. Any $S$ 
has a (minimal) generating set, with no common factors, such that 
every element of $S$ is a linear combination of generators with non-negative 
integer coefficients. For the above examples, the generators are 
$<3,4>$, $<4,6,9>$ and $<6,10,15>$.

The number of gaps $g(S)$ is called the genus, and the highest gap 
$F(S)$ is called the Frobenius number; they are respectively 3, 6, 15
and 5, 11, 29 in the above examples. $S$ is called a {\it symmetric}
numerical semigroup if $F(S)$ is odd, and for such $S$ it is a theorem that 
$F(S) + 1 = 2g(S)$. The examples above are all symmetric. A symmetric 
numerical semigroup $S$ has the nice property
that for each pair of integers $n$ and $F(S) - n$, with $0 \le n \le F(S)$, 
just one of the pair is in $S$ and the other is a gap. This makes it
easy to reconstruct the allowed spins from the first few. For example,
for $O_h$, if one remembers that the list starts with 0, 4 and 6 and that
there are no gaps from 12 onwards, then the only further gaps are at
$11 - 0 = 11$ and $11 - 4 = 7$. Similarly, there are only four gaps
beyond the element 15 in the $Y_h$ list, at 17, 19, 23 and 29.  

%\vfill
%\newpage

\section*{Appendix C: Table of Calcium-40 states}

Table 2 lists the experimentally observed isospin 0 states of 
Calcium-40, and their assignment to tetrahedral rovibrational bands 
in our model. Nearly all the states below 8
MeV are listed and assigned to bands, although a few dubious
states and some with completely undetermined spins are omitted. Between
8 MeV and just above 22 MeV, almost all states with spins greater than 6 are 
assigned to bands, as well as a few states with spins 6 or less. 
The first two columns in the table show the energy and spin/parity 
of each state as determined experimentally \cite{ENSDF}. Energies are
given to four significant figures here, although generally they are 
known to greater accuracy. Multiple spin/parity values and
brackets indicate uncertainty, or considerable theoretical input. The
third column is the spin/parity we have chosen from the experimentally
allowed possibilities, to obtain the most convincing rotational
bands. Extrapolation from better determined states point to gaps in
our bands, each with a definite spin/parity and a narrow energy 
range, and our choices are made to fill these. We have made 21 
choices, so some may not be correct. The fourth column shows the 
assigned band for the state.   

\begin{tabularx}{\linewidth}{ l l | l l } 
	
\hline \hline
Energy & $J^P$  & $J^P$ chosen & Band \\ \hline

0.0 & $0^+$ & $0^+$ & ${\rm g.s.}$ \\
3.353 & $0^+$ & $0^+$ & ${\rm A}$ \\
3.737 & $3^-$ & $3^-$ & ${\rm g.s.}$ \\
3.904 &	$2^+$ & $2^+$ & ${\rm E}$ \\	
4.491 &	$5^-$ & $5^-$ & ${\rm E}$ \\
5.212 &	$0^+$ & $0^+$ & ${\rm E}^2$ \\
5.249 &	$2^+$ & $2^+$ & ${\rm E}^2$ \\
5.279 &	$4^+$ & $4^+$ & ${\rm E / g.s.}$ \\
5.614 &	$4^-$ & $4^-$ & ${\rm E}$ \\
5.629 &	$2^+$ & $2^+$ & ${\rm F}$ \\
5.903 &	$1^-$ & $1^-$ & ${\rm F}$ \\
6.025 &	$2^-$ & $2^-$ & ${\rm E}$ \\
6.030 &	$3^+$ & $3^+$ & ${\rm F}$ \\
6.285 &	$3^-$ & $3^-$ & ${\rm A}$ \\
6.508 &	$4^+$ & $4^+$ & ${\rm F}$ \\                
6.543 &	$4^+$ & $4^+$ & ${\rm E}^2$ \\
6.582 &	$3^-$ & $3^-$ & ${\rm F}$ \\
6.750 &	$2^-$ & $2^-$ & ${\rm E}^2$ \\
6.909 &	$2^+$ & $2^+$ & ${\rm A} \times {\rm E}$ \\
6.930 &	$6^+$ & $6^+$ & ${\rm E}$ \\
6.931 &	$3^-$ & $3^-$ & ${\rm E}^2$ \\
6.938 &	$(1^- \ {\rm to} \ 5^-)$ & $4^-$ & ${\rm F}$ \\
6.950 &	$1^-$ & $1^-$ & ${\rm E \times {\rm F}}$ \\
7.100 &	$(2^+)$ & $2^+$ & ${\rm E} \times {\rm F}$ \\
7.113 &	$1^-$ & $1^-$ & ${\rm A} \times {\rm F}$ \\
7.114 &	$4^-$ & $4^-$ & ${\rm E}^2$ \\
7.239 &	$(3^-,4,5^-)$ & $5^-$ & ${\rm F}$ \\
7.278 &	$(2,3)^+$ & $3^+$ & ${\rm E} \times {\rm F}$ \\
7.301 &	$0^+$ & $0^+$ & ${\rm A}^2$ \\
7.397 &	$(5^+)$ & $5^+$ & ${\rm E}$ \\
7.446 &	$3^+,4^+$ & $3^+$ & ${\rm E} \times {\rm F}$ \\
7.466 &	$2^+$ & $2^+$ & ${\rm A} \times {\rm F}$ \\
7.532 &	$2^-$ & $2^-$ & ${\rm A} \times {\rm E}$ \\
7.561 &	$4^+$ & $4^+$ & ${\rm E}^2$ \\
7.623 &	$(2^-,3,4^+)$ & $4^+$ & ${\rm A}$ \\
7.677 &	$(6^+)$ & $6^+$ & ${\rm E}^2$ \\
7.702 &	$0^+$ & $0^+$ & ${\rm E}^3$ \\
7.769 &	$(3,4,5^-)$ & $5^-$ & ${\rm F}$ \\
7.815 &	$0^+$ & $0^+$ & ${\rm A} \times {\rm E}^2$ \\
7.870 &	$3^-$ & $3^-$ & ${\rm E} \times {\rm F}$ \\
7.872 &	$2^+$ & $2^+$ & ${\rm E}^3$ \\
7.928 &	$4^+$ & $4^+$ & ${\rm A} \times {\rm E}$ \\
7.972 &	$(\le 3)^-$ & $2^-$ & ${\rm E} \times {\rm F}$ \\
7.974 &	$(6^+)$ & $6^+$ & ${\rm E}^2$ \\
7.977 &	$2^+$ & $2^+$ & ${\rm A} \times {\rm E}^2$ \\
8.100 &	$8^+$ & $8^+$ & ${\rm E}$ \\
8.135 &	$(3^-)$ & $3^-$ & ${\rm E} \times {\rm F}$ \\		
8.187 &	$(3,4,5^-)$ & $5^-$ & ${\rm E}^2$ \\
8.338 &	$(2^+,3,4)$ & $3^+$ & ${\rm A} \times {\rm F}$ \\
8.359 &	$(0,1,2)^-$ & $0^-$ & ${\rm E}^3$ \\
8.364 &	$(3^- \ {\rm to} \ 7^-)$ & $3^-$ & ${\rm A} \times {\rm F}$ \\		
8.374 &	$4^+$ & $4^+$ & ${\rm E} \times {\rm F}$ \\
8.484 &	$(1^-,2^-,3^-)$ & $2^-$ & ${\rm E}^3$ \\
8.678 &	$4^+$ & $4^+$ & ${\rm E} \times {\rm F}$ \\
8.701 &	$(6^-)$ & $6^-$ & ${\rm E}$ \\
8.764 &	$3^-$ & $3^-$ & ${\rm E}^3$ \\
8.851 &	$6^-,7^-,8^-$ & $6^-$ & ${\rm E}^2$ \\		
8.936 &	$(7^+)$ & $7^+$ & ${\rm E}$ \\
8.978 &	$5^+,6^+,7^+$ & $6^+$ & ${\rm g.s.}$ \\
9.032 &	$4^-$ & $4^-$ & ${\rm E} \times {\rm F}$ \\
9.033 &	$(7^-)$ & $7^-$ & ${\rm E}$ \\		
9.246 &	$(7^-)$ & $7^-$ & ${\rm F}$ \\
9.305 &	$(8^+)$ & $8^+$ & ${\rm E}$ \\
9.429 &	$(3,4)^-$ & $4^-$ & ${\rm E} \times {\rm F}$ \\
9.454 &	$3^-$ & $3^-$ & ${\rm A}^2$ \\
9.811 &	$(3^-,4^-,5^-)$ & $5^-$ & ${\rm E} \times {\rm F}$ \\
9.853 &	$(8^+)$ & $8^+$ & ${\rm E}^2$ \\
9.860 &	$4^-,5^-,6^-$ & $6^-$ & ${\rm g.s.}$ \\		
9.869 &	$1^+,2^+$ & $1^+$ & ${\rm E} \times {\rm F}$ \\
9.921 &	$(3^-,4^-,5^-)$ & $4^-$ & ${\rm A} \times {\rm E}$ \\
10.05 & $(3^- \ {\rm to} \ 7^-)$ & $7^-$ & ${\rm E}^2$ \\
10.15 &	$(3^-,4^+,5^-)$ & $5^-$ & ${\rm E} \times {\rm F}$ \\
10.27 & $3^+,4^+,5^+$ & $5^+$ & ${\rm A} \times {\rm E}$ \\
10.47 & $(8^-)$ & $8^-$ & ${\rm E}$ \\
10.82 &	$3^+$ & $3^+$ & ${\rm E}^3$ \\
10.89 & $(9^-)$ & $9^-$ & ${\rm E}$ \\
11.00 & $(10^+)$ & $10^+$ & ${\rm E}$ \\
11.37 & $(5^-)$ & $5^-$ & ${\rm A} \times {\rm E}$ \\
11.69 & $(10^+)$ & $10^+$ & ${\rm E}$ \\
11.69 & $7^-$ & $7^-$ & ${\rm g.s.}$ \\
11.71 & $(9^+)$ & $9^+$ & ${\rm E}$ \\
12.33 & $(10^+)$ & $10^+$ & ${\rm E}^2$ \\
12.59 & $(10^+)$ & $10^+$ & ${\rm E}^2$ \\
12.92 & $(11^-)$ & $11^-$ & ${\rm E}$ \\
13.12 & $(12^+)$ & $12^+$ & ${\rm E}$ \\
13.19 & $(10^-)$ & $10^-$ & ${\rm E}$ \\
13.54 & $(11^+)$ & $11^+$ & ${\rm E}$ \\
13.62 & $6^+$ & $6^+$ & ${\rm A}$ \\
14.23 & $(12^+)$ & $12^+$ & ${\rm E}$ \\
14.87 & $(9^-)$ & $9^-$ & ${\rm g.s.}$ \\
15.15 & $(13^+)$ & $13^+$ & ${\rm E}$ \\
15.27 & $(12^+)$ & $12^+$ & ${\rm E}^2$ \\
15.31 & $(13^-)$ & $13^-$ & ${\rm E}$ \\
15.75 & $(12^+)$ & $12^+$ & ${\rm E}^2$ \\
16.53 & $(14^+)$ & $14^+$ & ${\rm E}$ \\
16.58 & $(13^+)$ & $13^+$ & ${\rm E}$ \\
17.70 & $(14^+)$ & $14^+$ & ${\rm E}$ \\
18.05 & $(14^+)$ & $14^+$ & ${\rm E}$ \\
18.21 & $(15^-)$ & $15^-$ & ${\rm E}$ \\
18.50 & $(14^+)$ & $14^+$ & ${\rm E}^2$ \\
18.72 & $(14^+)$ & $14^+$ & ${\rm E}^2$ \\
19.20 & $(15^+)$ & $15^+$ & ${\rm E}$ \\
20.58 & $(16^+)$ & $16^+$ & ${\rm E}$ \\
22.06 & $(16^+)$ & $16^+$ & ${\rm E}^2$ \\

\hline
		
\caption{Energy, spin/parity and rovibrational band assignments for 
states of Calcium-40.}

\end{tabularx}


\vspace{7mm}



\begin{thebibliography}{99}

\bibitem{Wheel} J. A. Wheeler,
Molecular viewpoints in nuclear structure,
{\it Phys. Rev. }{\bf 52} (1937) 1083.

\bibitem{Wef} W. Wefelmeier,
Ein geometrisches Modell des Atomkerns,
{\it Zeit. f. Phys. }{\bf A107} (1937) 332.

\bibitem{HT} L. R. Hafstad and E. Teller, 
The alpha-particle model of the nucleus,
{\it Phys. Rev. }{\bf 54} (1938) 681.

\bibitem{Lez} K. Lezuo, 
Ground state rotational bands in $^{16}{\rm O}$, $^{40}{\rm Ca}$ and 
$^{208}{\rm Pb}$?,
{\it Z. Naturforsch. }{\bf 30a} (1975) 158.

\bibitem{Heu} A. Heusler, Identification of rotating and vibrating 
tetrahedrons in the heavy nucleus $^{208}$Pb, 
{\it Eur. Phys. J. }{\bf A53} (2017) 215.

\bibitem{Den} D. M. Dennison,
Energy levels of the O$^{16}$ nucleus,
{\it Phys. Rev. }{\bf 96} (1954) 378.

\bibitem{Rob} D. Robson,
Evidence for the tetrahedral nature of $^{16}{\rm O}$,
{\it Phys. Rev. Lett. }{\bf 42} (1979) 876.

\bibitem{BI} R. Bijker and F. Iachello,
Evidence for tetrahedral symmetry in $^{16}$O,
{\it Phys. Rev. Lett. }{\bf 112} (2014) 152501.

\bibitem{HKM1} C. J. Halcrow, C. King and N. S. Manton,
Dynamical $\alpha$-cluster model of $^{16}{\rm O}$,
{\it Phys. Rev. }{\bf C95} (2017) 031303(R). 

\bibitem{HKM2} C. J. Halcrow, C. King and N. S. Manton,
Oxygen-16 spectrum from tetrahedral vibrations and their rotational
excitations,
{\it Int. J. Mod. Phys. }{\bf E28} (2019) 1950026.

\bibitem{Sky} T. H. R. Skyrme, 
A non-linear field theory,
\textit{Proc. Roy. Soc. Lond. }{\bf A260} (1961) 127.

\bibitem{BMS} R. A. Battye, N. S. Manton and P. M. Sutcliffe,
Skyrmions and the $\alpha$-particle model of nuclei,
{\it Proc. Roy. Soc. Lond. }{\bf A463} (2007) 261.

\bibitem{Her} G. Herzberg,
{\it Molecular Spectra and Molecular Structure: II. Infrared and Raman
  Spectra of Polyatomic Molecules},
Van Nostrand, Princeton NJ, 1945.

\bibitem{TWC} D. R. Tilley, H. R. Weller and C. M. Cheves,
Energy levels of light nuclei A = 16 -- 17,
{\it Nucl. Phys. }{\bf A564} (1993) 1.

\bibitem{Suh} J. Suhonen,
{\it From Nucleons to Nucleus},
Springer, Berlin Heidelberg, 2007.

\bibitem{Lau} P. H. C. Lau, unpublished.

\bibitem{LM} P. H. C. Lau and N. S. Manton,
Quantization of $T_d$- and $O_h$-symmetric Skyrmions,
{\it Phys. Rev. }{\bf D89} (2014) 125012.

\bibitem{Man} N. S. Manton,
Lightly Bound Skyrmions, Tetrahedra and Magic Numbers, 
arXiv:1707.04073 (2017).

\bibitem{HMR} C. J. Halcrow, N. S. Manton and J. I. Rawlinson,
Quantized Skyrmions from SU(4) Weight Diagrams, 
{\it Phys. Rev. }{\bf C97} (2018) 034307.

\bibitem{GHS} M. Gillard, D. Harland and M. Speight,
Skyrmions with low binding energies,
{\it Nucl. Phys. }{\bf B895} (2015) 272.

\bibitem{GHK} M. Gillard, D. Harland, E. Kirk, B. Maybee and M. Speight,
A point particle model of lightly bound Skyrmions,
{\it Nucl. Phys. }{\bf B917} (2017) 286.

\bibitem{Ide} E. Ideguchi et al.,
Superdeformation in the doubly magic nucleus $^{40}_{20}{\rm Ca}_{20}$,
{\it Phys. Rev. Lett. }{\bf 87} (2001) 222501.

\bibitem{Tor} S. Torilov et al.,
Spectroscopy of $^{40}{\rm Ca}$ and negative-parity bands,
{\it Eur. Phys. J. }{\bf A19} (2004) 307.

\bibitem{ENSDF} Evaluated Nuclear Structure Data File,

https://www.nndc.bnl.gov/ensdf/index.jsp .

\bibitem{LL} L. D. Landau and E. M. Lifshitz,
{\it Quantum Mechanics -- Course of Theoretical Physics Vol. 3 (3rd ed.)},
Butterworth--Heinemann, Oxford, 1977.

\bibitem{NumSem} https://en.wikipedia.org/wiki/Numerical\_semigroup.

\bibitem{BJ} P. R. Bunker and P. Jensen, 
Spherical top molecules and the molecular symmetry group,
{\it Mol. Phys. }{\bf 97} (1999) 255.

\end{thebibliography}
\end{document}